\documentstyle[11pt,amsfonts,amssymb,amsmath,amscd]{article}
\begin{document}
\title{Noncommutative Geometry, Hodge Theorem and Holography}
\author{Ioannis P. ZOIS
\\
School of Natural Sciences, Department of Mathematics
\\
The American College of Greece, Deree College
\\
 6 Gravias Sstreet, GR-153 42
\\
Aghia Paraskevi, Athens, Greece
\\
e-mail: i.zois@exeter.oxon.gov}

\newtheorem{thm}{Theorem}
\newtheorem{defn}{Definition}
\newtheorem{prop}{Proposition}
\newtheorem{lem}{Lemma}
\newtheorem{cor}{Corollary}
\newtheorem{rem}{Remark}
\newtheorem{ex}{Example}

\newcommand{\Rat}{\mathbb Q}
\newcommand{\Real}{\mathbb R}
\newcommand{\RR}{\Real}
\newcommand{\Rh}{\hat{\Real}}
\newcommand{\Nat}{\mathbb N}
\newcommand{\Complex}{\mathbb C}
\newcommand{\HH}{\mathbb H_3}
\newcommand{\CC}{\Complex}
\newcommand{\Z}{\mathbb Z}

\newcommand{\Ea}{{\mathcal E}}
\newcommand{\Ta}{{\mathcal A}}
\newcommand{\Aa}{{\Ta_\infty}}
\newcommand{\Eb}{{C^*(D_1,D_2,X_2,\Omega)}}
\newcommand{\Tb}{{C^*(D_1,D_2,\hat\Omega)}}
\newcommand{\Ab}{{C^*(D_1,D_2,\Omega)}}
\newcommand{\aco}{\idl}
\newcommand{\aca}{\beta^{\|}}
\newcommand{\acb}{\beta^\perp}
\newcommand{\acc}{{\hat\tau}}

\newcommand{\Uu}{{\cal U}}
\newcommand{\Dd}{{\mathcal D}}
\newcommand{\Oo}{{\mathcal O}}
\renewcommand{\H}{{\mathcal H}}
\newcommand{\NN}{{\bf N}}
\newcommand{\ZZ}{{\bf Z}}
\newcommand{\Pp}{{\mathcal P}}
\newcommand{\Zz}{{\mathcal Z}}
\newcommand{\PP}{{\bf P}}
\newcommand{\LL}{\Lambda}
\newcommand{\LLL}{\Lambda\cup \infty}
\newcommand{\EE}{{\bf E}}
\newcommand{\Bb}{{\mathcal B}}
\newcommand{\Ww}{{\mathcal W}}
\newcommand{\Ss}{{\mathcal S}}

\newcommand{\Tr}{\mbox{\rm Tr}}  
\newcommand{\TV}{{\mathcal T}}
\newcommand{\TVh}{\hat{{\mathcal T}}}
\newcommand{\tr}{\mbox{tr}}

\newcommand{\Rr}{{\mathcal R}}
\newcommand{\Nn}{{\mathcal N}}
\newcommand{\Cc}{{\mathcal C}}
\newcommand{\Jj}{{\mathcal J}}
\newcommand{\Ff}{{\mathcal F}}
\newcommand{\Ll}{{\mathcal L}}
\newcommand{\id}{{\mbox{\rm id}}}
\newcommand{\idl}{{\mbox{\rm\tiny id}}}
\newcommand{\eva}{{\mbox{\rm ev}}}
\newcommand{\eval}{{\mbox{\rm\tiny ev}}}

\def\essinf{\mathop{\rm ess\,inf}}
\def\esssup{\mathop{\rm ess\,sup}}

\newcommand{\bew}{{\bf Proof:}}
\newcommand{\eb}{\hfill $\Box$}

\newcommand{\x}{{\vec x}}
\newcommand{\y}{{\vec y}}
\newcommand{\n}{{\vec n}}
\renewcommand{\a}{{\vec a}}
\newcommand{\hull}{\Sigma}
\newcommand{\om}{\omega}
\newcommand{\oh}{{\hat{\om}}}

\newcommand{\Af}{{\Aa}_0}
\newcommand{\Tf}{{\Ta}_0}
\newcommand{\Ef}{{\Ea}_0}
\renewcommand{\H}{{\mathcal H}}

\newcommand{\bs}{\bigskip}
\newcommand{\ms}{\medskip}

\newcommand{\erz}[1]{\langle{#1}\rangle}
\newcommand{\pair}[2]{\erz{#1,#2}}
\newcommand{\diag}[2]{\mbox{\rm diag}(#1,#2)}
\newcommand{\tOmega}{{\Omega^s}}
\newcommand{\td}{{d^s}}
\newcommand{\tint}{{\int^s}}
\newcommand{\ttau}{{\tilde\tau}}
\newcommand{\talpha}{{\tilde\alpha}}
\newcommand{\tS}{{\tilde S}}
\newcommand{\ual}{{\underline{\alpha}}}
\newcommand{\hotimes}{{\hat\otimes}}
\newcommand{\K}{{\mathcal K}}
\newcommand{\Ypsilon}{{\Theta}}
\newcommand{\CA}{$C^*$-algebra}
\newcommand{\CF}{$C^*$-field}
\newcommand{\G}{\mathcal G}
\newcommand{\im}{\mbox{\rm im\,}}
\newcommand{\rk}{\mbox{\rm rk\,}}
\newcommand{\cp}{{\rtimes}}
\newcommand{\del}{{\bf \delta}}

\renewcommand{\a}{{\vec a}}
\newcommand{\at}{{\bf \tilde a}}
\newcommand{\cy}{\Phi}
\newcommand{\co}{\cy_\a}
\newcommand{\tco}{\cy_{\at}}

\newcommand{\ttimes}{{\tilde \rtimes}}
\renewcommand{\ss}{{\mathcal R}}
\renewcommand{\Cc}{{\mathcal C}}
\newcommand{\Ri}{(\RR\cup\infty)}
\newcommand{\Rd}{\hat\RR}
\newcommand{\dom}{\mbox{\rm dom}}
\newcommand{\fin}{\mbox{\rm\tiny fin}}

\newcommand{\enn}{\mu}
\newcommand{\ch}{\mbox{\rm ch}}
\newcommand{\supp}{\mbox{\rm supp}}
\newcommand{\chd}{\ch}
\newcommand{\Oh}{\hat\hull}

\newcommand{\hsp}{\RR^{d-1}\times\RR^{\leq 0}}
\newcommand{\IDS}{IDS}
\newcommand{\sv}[2]{\left(\begin{array}{c} #1 \\ #2 \end{array}\right)}

\bibliographystyle{amsalpha}

\maketitle

\begin{abstract}
Some time ago we presented an article (which was in fact the outline of a research programme) in which we argued for the need to develop a nonommutative version of topological quantum field theories (NCTQFT for short).  Recent work by C.J. Hogan et all, has demonstrated the possibility to get experimental verification of holography; if this comes true, then that would indicate that quantum gravity is indeed a TQFT. On the other hand there is accumulating evidence that the underlying geometry of spacetime is a noncommutative (abreviated to nc in the sequel) space, hence if one wants a unified theory of all physical interactions including gravity that would mean that the right framework would be NCTQFT. Therefore there is strong motivation for this research. Towards this goal we present a modest achievement which is a nc version of Hodge Theorem and the definition of the nc free bosonic propagator.


\end{abstract}

\section{Introduction}

Some years ago in \cite{z1} we argued that one should try to construct noncommutative topological quantum field theories (NCTQFT for short) as a natural extension of the usual TQFT's. The motivation came from accumulating evidence that the fundamental geometry of spacetime is not Riemannian geomtry but noncommutative geometry (NCG for short). Moreover during this current year (2009) there have been some remarkable indications that the holography principle for quantum gravity (whose mathematical expression is that quantum gravity is a TQFT) might be verified experimentally. Thus our interest in the construction of NCTQFT's increased since this should be the right framework for a quantum theory of all known physical interactions including gravity. Towards this goal we present a NC version of Hodge Theorem and the construction of the NC free bosonic propagator.\\

We organise this article as follows: First we briefly review the  holography principle for quantum gravity and its relation to TQFT's. Next we review the recent experimental evidence towards the verification of the holography principle related to the work of C.J. Hogan (et all) and the experiment GEO 600. Then we review the evidence indicating that the underlying geometry of spacetime is in fact NCG and  we present the NC Hodge Theorem and the construction of the nc free bosonic propagator.\\

\section{Holography and TQFT}

Back in 1993, G. 't Hooft in \cite{thooft} introduced the principle of holography (his original term was "dimensional reduction", the term holography was introduced later by L. Susskind) which states the following:\\

\emph{"Quantum Gravity (QG) on a $(d+1)$-manifold with boundary is equivalent to a Conformal Field Theory (CFT) on the boundary (which obviously is a $d$-manifold). Moreover on each fundamental $d$-hypersurface with area $l_{P}^d$ (where $l_P=10^{-35}$m is the Planck length) on the boundary, there corresponds one degree of freedom of the theory on the bulk".}\\

The M-Theoretic version of the holography principle is the so-called Maldacena conjecture relating gravity on anti se Sitter spaces with CFT on the boundary using S-duality.\\

The "lost" dimension corresponds to \textsl{rescaling} which justifies the use of conformal symmetry.\\

The mathematical interpretation of holography is the statement that\\ 

\emph{"QG must be a TQFT"}.\\

The motivation behind holography is clearly the famous and very counter-intuitive formula of Bekenstein-Hawking which states that the entropy of a black hole is proportional to the \textsl{area} of its event horizon (and NOT the volume).\\

The formal definition of a TQFT was given by M.F. Atiyah (for example one can see \cite{atiyah1} and \cite{atiyah2}). A more heuristic definition would be that \textsl{a TQFT is a theory whose partition function is a (numerical) topological invariant} (on a closed manifold).\\
 
It is straightforward for one to realise that this should be the case for QG: The action for gravity on a $4$-dim (pseudo)Riemannian manifold $(M^4,g)$ is the Einstein-Hilbert action

$$S(M^4,g)=\frac{1}{2K}\int_{M^4}RdV$$
where

$$K=\frac{8\pi G}{c^4},$$
is a constant,
$$dV=\sqrt{-g}d^4 x$$
is the fundamental volume form of the $4$-mnifold and $R=R(g)$ is the scalar curvature which depends only on the (pseudo)Riemannian metric $g$.\\

The standard recipe to quantize a theory is to write down the partition function which in this case would be

$$Z[M^4]=N\int e^{iS}Dg$$

where $N$ is a normalisation constant.\\ 

It is obvious that if we managed to get a finite result out of the above path integral, the corresponding partition function would depend only on the topology of the manifold since the metrics would have been integrated out (the Einstein-Hilbert action depends on 2 data/"variables", topology and metric, if we plug that in the path integral and integrate the metrics, only the topology will survive thus the result should be a topological inavariant).\\

Holography is closely related to the \emph{Deligne Conjecture} in Algebraic Geometry by the use of \emph{operads}, see \cite{z2}.\\

\section{Holography and GEO 600}

Although the quantization of gravity appears to be the most important open problem in physics today, it is not even sure that such a theory exists at all. The reason for this is because quantum theory, as we know it for almost a century, is based on the well-known \emph{particle-wave duality} whose mathematical expression is the uncertainty principle. \emph{Until today, for the case of the gravitational interactions, there is no experimental evidence neiher for the existence of gravitational waves nor for the existence of the gravitons} (the "carriers" of the gravitational interaction). Thus for gravity, both sides of the particle-wave duality, which is the cornerstone of quantum theory, are currently missing!\\

Yet despite the above, most physicists believe that QG exists. The basic arguments are the following:\\

{\bf 1.} QG should exist due to mathematical consistency of Einstein's classical equations (no cosmological constant)
$$G_{ab}:=R_{ab}-\frac{1}{2}Rg_{ab}=8\pi GT_{ab}.$$
The RHS contains the energy-momentum tensor due to mass and energy of all other fields present except gravity, this consists of ordinary matter (hence quarks, leptons, neutrinos and possibly the Higgs) along with the energy of the electroweak and strong fields, all these fields are quantized. Thus for mathematical consistency the LHS (the Einstein tensor which incodes the geometry) should also be quantized.\\

{\bf 2.} We want to remove the singularities appearing form the Hawking-Penrose singularity theorems.\\

{\bf 3.} We want compatibility between the physics of the small scales with the large scale physics (this includes issues like the arrow of time and the 3rd law of thermodynamics, the entropy of black holes, the Hawking radiation, the no-boundary proposal of Hartle and Hawking etc).\\

{\bf 4.} Various philosophical and aesthetic arguments.\\

Gravitons are extremely difficult to detect because they interact very weakly with all other particles. Yet most physicists are convinced that gravitational waves exist. Currently there are 3-4 experiments around the globe aiming at detecting them, one of which is the Anglo-German experiment GEO 600. The idea is simple: Since tidal forces stretch the vertical direction and squeeze the horizontal, the apparatus cinsists of 2 ultra sensitive huge rulers forming the sides of a right angle and measuring relative variations of distances travelled by emitted photons in the two perpendicular directions. Thus basically the apparatus is a gigantic interferometer. For several months an inexplicable background noise at about 700Hz puzzled the GEO 600 researchers at Hanover.\\

The only explanation (up to now) for this noise was given by C.J. Hogan in \cite{hogan} based on holography. The starting point was the following idea: If we assume that the universe at some time instant $t$ has a certain radius $R$, then one should have
$$n=\frac{4\pi R^2}{l_P^2}$$
degrees of freedom on the boundary 2-sphere. Holography then states that one should have the same number of degrees of freedon in the bulk, hence if we divide the volume of the 3-sphere by the volume of an  elementary cube with edge $x$ (and volume $x^3$), then one should have 
$$n=\frac{4\pi R^2}{l_{P}^2}=\frac{(4/3) \pi R^3}{x^3}.$$
Solving the above equation one finds that $x\sim 10^{-16}m$, namely if holography is true, we should be able to detect various phenomena related to it with current technology (recall that the size of a proton is $\sim 10^{-15}m$). (One tricky point here is the radius of the universe since the universe expands rapidly due to dark energy; Hogan assumed that the radius of the universe is the distance from the "boundary" from beyond which light has not had time to reach us in the $13.7$ billion year life span of the universe). Then, in that article (and some more that followed),
a particular form for the quantum indeterminacy of relative spacetime position of events was derived
from the limits of measurement possible with Planck wavelength radiation. The indeterminacy
predicted fluctuations from a classically defined geometry in the form of  holographic noise whose
spatial character, absolute normalization, and spectrum were predicted with no parameters. The
noise had a distinctive transverse spatial shear signature, and a flat power spectral density given by
the Planck time. An interferometer signal displays noise due to the uncertainty of relative positions
of reflection events. The noise corresponds to an accumulation of phase offset  with time that mimics
a random walk of those optical elements that change the orientation of a wavefront. It only appears
in measurements that compare transverse positions  and does not appear at all in purely radial
position measurements. A lower bound on holographic noise followed from a covariant upper bound
on gravitational entropy. And then, perhaps to everyone's surprise,\emph{the predicted holographic noise spectrum was eventually estimated to be comparable to the measured noise in the currently operating interferometer GEO600}.\\

We definetely need extra verifications before coming to any concrete conclusions (one can compare the situation however with the discovery of the cosmic microwave background radiation about half a century ago).\\

\section{Noncommutative Geometry}

The previous discussion was about pure gravity (namely no other fields present except the gravitational filed). During the last decade however there is increasing evidence that the fundamental geometry of spacetime is not (pseudo)Riemannian geometry but Noncommutative geometry (ncg). This evidence comes both from theories living in more than 4 dimensions (like superstring theory, M-Theory etc) but also from plain 4-dimensional theories.\\

More specifically, back in 1998 Connes (et all) in \cite{connes1} proved that type IIB superstrings admit aditional compactifications on noncommutative tori. The proof is indirect: First the authors prove the result for the IKKT matrix model and then they prove that the IKKT matrix model is equivalent to type IIB superstrings when the size of matrices tends to infinity.\\

The following year, Seiberg and Witten in \cite{witten} proved a similar result for type I superstrings on D-branes in the presence of a constant B-filed. Various generalisations appeared later.\\

A couple of years ago, Connes et all in an impressive article (building on and sharpening previous work, see \cite{connes2}), managed to indicate that the standard model Lagrangian density (this lives obviously in 4 dimensions and contains all fields which have been verified experimentally like gravity, quarks, leptons, neutrinos, Yang-Mills fields for electroweak and strong forces, Yukawa couplings, Higgs field etc), can be thought off as the fundamental K-Homology class of a noncommutative space which is the product of a spin 4-manifold with a discrete space of metric dimension 0 and KO-dimension 6 mod 8.\\

Along the same lines one could mention a similar result concerning 11 dimensional supergravity (see \cite{zois1}).\\

\emph{All the above seem to indicate that should one want to seek for a quantum theory of all physical interactions including gravity, the correct framework would be NCTQFT, where holography (and quantum gravity) is responsible for the "topological" character of this would-be theory and the presence of other fields are responsible for noncommutativity}.\\

Towards this ambitious goal, some modest progress has been achieved: An analogue of the Hodge theorem can be stated for NCG and as an immediate  corollary a nc free bosonic propagator can be constructed.\\

Let us briefly recall that the "clasical" Hodge Theorem states that on every smooth, compact, Riemannian manifold (also assumed oriented), each de Rham cohomology class has a unique harmonic representative (namely the Laplace operator vanishes). As a consequence, every closed form can be written as the sum of an exact form plus a harmonic form and moreover every form can be written as the sum of a harmonic form plus an exact form plus a coexact form.\\

We follow Quillen (\cite{quillen1}, \cite{quillen2}): Let $A$ be a complex, unital associative algebra and let
$$\Omega ^{n}A:=A\otimes _{\mathbb{C}}\bar{A}^{\otimes n},$$
for $n>0$, where
$$\bar{A}=A/\mathbb{C}$$
whereas
$$\Omega ^{n}A=0, n<0$$
and
$$\Omega ^{0}A=A.$$
Hence we get an identification
$$a_0da_1...da_n\leftrightarrow (a_0,a_1,...,a_n).$$
Then we also define
$$\Omega A=\oplus _n\Omega ^nA$$
which is the graded algebra (GA) of noncommutative differential forms over $A$, the multiplication being defined via
$$(a_0,a_1,...,a_n)(a_{n+1},a_{n+2},...,a_{k})=\sum_{i=0}^{k}(-1)^{k-i}(a_0,...,a_{i}a_{i+1},...,a_k)$$
for $k>n$. Moreover we define the differential $d:\Omega ^nA\rightarrow\Omega ^{n+1}A$ as follows:
$$d(a_0da_1...da_n)=da_0da_1...da_n$$
or in an equivalent notation
$$d(a_0,a_1,...,a_n)=(1,a_0,a_1,...,a_n)$$
and hence
$$d\Omega ^nA\simeq\bar{A}^{\otimes n+1}$$
for $n\geq 0$. Thus $(\Omega A,d)$ becomes a DGA.\\

On $\Omega A$, we can also define the Hochschild differential
$b:\Omega ^{n}A\rightarrow\Omega ^{n-1}A$
given by 
$$b(a_0,a_1,...,a_n)=\sum _{j=0}^{n-1}(-1)^{j}(a_0,a_1,...,a_{j}a_{j+1},...,a_n)+(-1)^{n}(a_na_1,a_2,...,a_{n-1}).$$
Thus one has that
$$b(\omega da)=(-1)^{|\omega |}(\omega a-a\omega )=(-1)^{|\omega |}[\omega ,a]$$
and
$$b(a)=0,$$ 
where $|\omega |$ denotes the degree of the differential form $\omega$.\\

One also has the Karoubi operator (see \cite{kar}) which is a degree zero operator on $\Omega A$ given by
$$k:\Omega ^{n}A\rightarrow\Omega ^{n}A$$
where
$$k(\omega da)=(-1)^{|\omega |}(da)\omega $$
(for negative degrees it is given by the identity).\\

{\bf Lemma 1.} \textsl{One has the following relation:
$$bd+db=1-k.$$}
{\bf Proof:} One has
$$(bd+db)(\omega da)=b(d\omega da)+(-1)^{|\omega |}d[\omega ,a]=(-1)^{|\omega |+1}[d\omega ,a]+(-1)^{|\omega |}d[\omega ,a]=$$
$$=[\omega ,da]=\omega da - (-1)^{|\omega |}(da)\omega.$$
$\square$\\

An immediate corollary of the above is that $k$ commutes both with $d$ and $b$, namely
$$bk=kb$$
and
$$dk=kd.$$
The above also shows that $k$ is homotopic to the identity with respect to either of the differentials $b$ or $d$.\\

One can formally see $d$ and $b$ as adjoint to each other, playing the roles of $d$ and $d^*$ respectively on the de Rham complex of a Riemannian manifold and call
$$bd+db=1-k$$
the \emph{NC Laplacian}. The natural thing to do next is to examine the spectrum of the NC Laplacian and focus on the zero eigenvalue. We have the following result:\\

{\bf Proposition 1.} \textsl{On $\Omega A$ one has the \emph{harmonic decomposition}
$$\Omega A=Ker(1-k)^2\oplus Im(1-k)^2,$$
where the generalised nullspace $Ker(1-k)^2$ is analogous to the space of harmonic forms}.\\

We can define the \emph{harmonic projection} $P$ to be the projection operator which is one on the first term of the harmonic decomposition and zero on the second; it is the spectral projection for $k$ associated to the eigenvalue 1. Hence the harmonic decomposition can be written
$$\Omega A=P\Omega A\oplus P^{\perp}\Omega A$$
where by definition
$$P^{\perp}=1-P$$
is the spectral projection of $k$ associated to the set of eigenvalues which are different from 1.\\

{\bf Proof:} The proof is based on the following technical Lemma:\\

{\bf Lemma 2.} \textsl{The Karoubi operator $k$ on $\Omega ^{n}A$ satisfies the polynomial relation
$$(k^n-1)(k^{n+1}-1)=0.$$}
{\bf Proof of Lemma 2:} We have
$$k(a_0da_1...da_n)=(-1)^{n-1}da_na_0da_1...da_{n-1}=$$
$$=(-1)^n a_nda_0...da_{n-1}+(-1)^{n-1}d(a_na_0)da_1...da_{n-1}$$
which in the second equivalent notation reads
$$k(a_0,a_1,...,a_n)=(-1)^n(a_n,a_0,...,a_{n-1})+(-1)^{n-1}(1,a_na_0,...,a_{n-1}).$$
Moreover
$$k(da_0da_1...da_n)=(-1)^nda_nda_0...da_{n-1}.$$  
In particular on $\Omega ^nA$ we have that $k^{n+1}d=d$.\\

Next we consider
$$k^j(a_0da_1...da_n)=(-1)^{j(n-1)}da_{n-j+1}...da_na_0da_1...da_{n-j}$$
for $0\leq j\leq n$. Hence
$$k^n(a_0da_1...da_n)=da_1...da_na_0=a_0da_1...da_n+[da_1...da_n,a_0]=$$
$$=a_0da_1...da_n+(-1)^nb(da_1...da_nda_0)$$
which yields
$$k^n=1+bk^nd$$
on $\Omega ^n A$. Then
$$k^{n+1}=k+bk^{n+1}d=k+bd$$
and using the definition of the NC Laplacian we get
$$k^{n+1}=1-db.$$
Thus from 
$$k^n=1+bk^nd$$
and
$$k^{n+1}=1-db$$
we obtain that $k$ on $\Omega ^n A$ satisfies the polynomial relation
$$(k^n-1)(k^{n+1}-1)=0.$$
$\square$\\
 
This polynomial relation implies that $k$ is invertible since the polynomial has constant term 1.\\

We returm to the proof of Proposition 1: Since an operator satisfies a polynomial equation, it gives rise to a direct sum decomposition into generalised eigenspaces corresponding to the distinct roots of the polynomial.\\

The roots of 
$$(k^n-1)(k^{n+1}-1)$$
are the $n$ different n-th roots of unity and the $n+1$ different roots of unity of order dividing $n+1$. Yet $n$ and $n+1$ are relatively prime which means that these two sets of roots have only $k=1$ in common. Hence 1 is a double root and and all other roots are simple.\\

Consequently $\Omega ^nA$ decomposes into the direct sum of the generalised eigenspace $Ker(1-k)^2$ corresponding to the eigenvalue $z=1$ and the ordinary eigenspaces $Ker(k-z)$ for each root of unity $z\neq 1$ of order dividing $n$ or $n+1$.\\

Combining the above $\forall n$ we obtain the following spectral decomposition with respect to $k$
$$\Omega A =Ker (1-k)^2 \oplus [\bigoplus _{z\neq 1}Ker (k-z )].$$
Lumping the eigenvalues $z\neq 1$ together we have
$$\Omega A=Ker(1-k)^2\oplus Im(1-k)^2$$
which completes the proof.\\

$\square$\\

Note however that the NC Laplacian
$$1-k=[b,d],$$
contrary to the Riemannian manifold situation, \textsl{is only nilpotent on the first factor (and invertible on the second)}. This defect can be cured by introducing the \emph{rescaled NC Laplacian}
$$L=[b,Nd]$$
where $N$ is the numbering operator (this is a degree zero operator which acting on forms gives the scalar multiple of the form by its degree). \emph{The rescaled NC Laplacian then vanishes on $P\Omega A$ (and is invertible on its complement)}.\\

On the complementary space $P^{\perp}\Omega A$ the NC Laplacain is invertible and hmotopic to zero with respect to either differential $b$ or $d$. Thus we can define the \emph{Green's operator} $G$ for the NC Laplacian which is equal to its inverse on $P^{\perp}\Omega A$, namely
$$G=(1-k)^{-1}$$
and $G=0$ on $P\Omega A$. This can be seen as the \emph{NC free bosonic propagator}.\\

As in the clasical Hodge theory, the complementary space $P^{\perp}\Omega A$ to the "harmonic fomrs" splits into subspaces of exact and coexact forms:\\

{\bf Proposition 2.} \textsl{One has}
$$P^{\perp}\Omega A=dP\Omega A\oplus bP\Omega A.$$

{\bf Proof:} This is a formal cosnequence of the identity
$$G(bd+db)=1$$
on $P^{\perp}\Omega A$ and the fact that $G$ commutes with both differentials $b$,$d$ (as one can check via a direct computation). Thus
$$(Gdb)d=G(bd+db)d=d$$
implies that $Gdb$ is a projection with image $dP^{\perp}\Omega A$. Similarly $Gbd$ is a projection with image $bP^{\perp}\Omega A$ and as these projections add to 1 we get the desired decomposition.\\

$\square$\\ 

Let us close this article with some comments: There are 5 basic TQFT's known up to now: The (2+1) Abelian Chern-Simons theory due to Albert Schwarcz, its non-Abelian generalisation (this is the so-called Jones-Witten theory), the (3+1) Donaldson-Floer -Witten theory, its dual (the so-called Seiberg-Witten theory) and the Kontsevich-Gromov-Witten theory (topological $\sigma$ models) and their generalisations.\\

The simplest of all is the Abelian Chern Simons theory where the Lagrangian density is given by the Abelian Chern-Simons 3-form
$$S=\int_{N^3}A\wedge dA$$
and the partition function is given by the following product of zeta-function regularised determinants of Laplacians
$$Z(N^3)=(Det_{\zeta}\Delta _{1})^{-1/4}(Det _{\zeta}\Delta _{0})^{3/4}.$$

\emph{The NC version of this should be obtained in a straightforward way for the case of a, say, noncommutative 3-sphere and using the NC Laplacian. We hope to be able to report on this elsewhere} (see \cite{quillen3} for Chern-Simons forms in NCG and \cite{connes3} and \cite{connes4} for noncommutative 3-spheres). Moreover it would be nice to see that the new NC topological invariant defined in \cite{z} will appear somewhere.\\

\end{document}